\documentclass[12pt]{iopart}

\usepackage{graphicx} 
\usepackage{iopams}
\begin{document}
\title{Microwave-mediated heat transport in a quantum dot attached to leads}
\author{Feng Chi$^1$,  Yonatan Dubi$^{2}$}
\address{$^1$ College of Engineering, Bohai University,
Jinzhou 121013, China\\ $^2$
Landa Laboratories, 3 Pekeris St., Rehovot 76702, Israel}
\ead{chifeng@semi.ac.cn}
\begin{abstract}
The thermoelectric effect in a quantum dot (QD) attached to two
leads in the presence of microwave fields is studied by using the
Keldysh nonequilibrium Green function technique. When the
microwave is applied only on the QD and in the linear-response
regime, the main peaks in the thermoelectric figure of merit and
the thermopower are found to decrease, with the emergence of a set
of photon-induced peaks. Under this condition the microwave field
can not generate heat current or electrical bias voltage.
Surprisingly, when the microwave field is applied only to one
(bright) lead and not to the other (dark) lead or the QD, heat
flows mostly from the dark to the bright lead, almost
irrespectively to the direction of the thermal gradient. We
attribute this effect to microwave-induced opening of additional
transport channels below the Fermi energy. The microwave field can
change both the magnitude and the sign of the electrical bias
voltage induced by the temperature gradient.

\end{abstract}

\pacs{73.21.La, 72.15.Jf, 73.50.Pz, 73.23.Hk}
\maketitle

\section{Introduction}
Owing to the possibility of high heat-voltage conversion
efficiency, thermoelectric phenomenon in nanoscale solid-state
materials has become an active research area in recent years[1,2].
In the linear response regime, i.e., the temperature gradient in
the device $\Delta T$ approaches to zero, the thermoelectric
efficiency is measured by the dimensionless thermoelectrical
figure of merit $ZT$, and is usually smaller than one in bulk
materials[3]. This hinders the wide usage of thermoelectric effect
in commercial applications. The reason behind the small $ZT$ in
bulk materials is the Wiedemann-Franz law[4], which can be
violated in nanoscale materials, for instance due to the Coulomb
blockade effect[5]. In the nonlinear response regime, i.e., finite
$\Delta T$ case, the thermoelectric energy conversion efficiency
is characterized by the bias voltage $\Delta V$ generated by
$\Delta T$[6]. In both of linear and nonlinear cases, the
thermoelectric efficiency in nanostructures may be much higher
than that in bulk materials because of the reduced
dimensionality[7-9], and more interestingly, depends on the
quantized energy levels that can be modulated by gate voltages.
Since the breakthroughs in experimental work[10-14], large figure
of merit exceeding one or higher and large $\Delta V$ were
frequently reported in superlattices, quantum wires, quantum dots,
and carbon nanotubes, etc. Theoretically, impacts of Coulomb
blockade[5-9,15-20], Kondo[21-24], and Fano[25,26] effects on the
thermoelectric properties were extensively investigated.
Applications such as nanoscale refrigeration[27], thermal
rectifier[18,28,29], thermal transistors[6], and thermal memory
and logic gates[30] were proposed.

The above-mentioned works mainly focused on the thermoelectric
efficiency under time-independent fields. Very recently, heat
transport with photons has also been studied. It was demonstrated
that heat can be conducted by photon radiation at very low
temperature when the phonons are frozen out[31]. Quantum circuits
sandwiched between two reservoirs and threaded by electromagnetic
fluctuations (photons) were proposed to operate as heat
transistors or electron coolers[32,33]. In a double-quantum-well
structure, it was shown that hotter electrons in one reservoir may
be replaced by the cooler ones in the other reservoir with the
help of photon by adjusting the wells' energy levels[34]. Electron
heating in two-dimensional electron system has also been
theoretically and experimentally investigated[35]. In addition,
Lau \emph{et al.} have calculated the properties of coherent
radiative thermal conductance in photonic crystals[36]. Moreover,
enhanced thermopower (Seebeck coefficient) induced by a
time-dependent gate voltage was proposed in a device of a quantum
dot (QD) coupled to two metal leads[37].

As a zero-dimensional system, QD is a promising thermoelectrical
material and was intensively studied in many of the above
mentioned works. Until now, photon-assisted heat transport in such
a device has seldom been addressed, although it has been studied
in the context of electron transport through QDs long time
ago[38,39]. In this paper, we study the photon-mediated
thermoelectric effect in a QD attached to two leads in both linear
($\Delta T \rightarrow 0$) and nonlinear (finite $\Delta T$)
regimes. Compared with previous studies on quantum
circuits[32,33], we deal with a relatively high temperature
situation. Our main results show that when the microwave field is
applied only on the QD, the magnitude of the figure of merit is
suppressed because of the photon-induced sub-channels in the dot.
When the microwave is applied only on one lead, large heat current
and electric bias are generated. We find that the heat current
generally flows from the dark lead into the bright one, except for
the situation in which the dot level is around the electron-hole
symmetry point. In the present paper, the contributions from
phonons are neglected for the sake of simplicity.
\section{Model and method}
The present system can be described by the following
Hamiltonian[39,40]:
\begin{eqnarray}
H(t)&=&\sum_{k,\sigma,\beta}\varepsilon_{k\beta}(t)c_{k\beta\sigma}^{\dag}c_{k\beta\sigma}+\sum_{\sigma}\varepsilon_{d}(t)d_{\sigma}^{\dag}d_{\sigma}+
+ Ud_{\uparrow}^\dag d_{\uparrow} d_{\downarrow}^\dag
d_{\downarrow}\nonumber\\
&+&\sum_{k,\sigma,\beta}(V_{\beta
d}c_{k\beta\sigma}^{\dag}d_{\sigma}+\mathrm{H.c.}),
\end{eqnarray}
\noindent where $c_{k\beta\sigma}^{\dag}$ $ (c_{k\beta\sigma})$ is
the creation (annihilation) operator of the electrons with
momentum $k$, spin $\sigma$ and energy $\varepsilon_{k\beta}(t)$
in the $\beta$ $(=L,R)$ lead;  $d_{\sigma}^{\dag}$ $ (d_{\sigma})$
creates (annihilates) an electron with spin $\sigma$, energy
$\varepsilon_d(t)$ in the QD. $U$ denotes the intradot Coulomb
interaction. The last term in Eq. (1) represents the dot-lead
coupling with coupling strength $V_{\beta d}$. The time-varying
energy spectra originate from the applied microwave fields and
take the forms of
$\varepsilon_{k\beta}(t)=\varepsilon_{k\beta}+\Delta_\beta\cos\omega
t$ and $\varepsilon_{d}(t)=\varepsilon_{d}+\Delta_d\cos\omega t$,
where $\varepsilon_{k\beta}$ and $\varepsilon_{d}$ are the
time-independent single electron energies without microwave field,
and $\Delta_{\beta/d}$, $\omega$ are respectively the microwave
strength and frequency applied on the leads or the dot[39,40].

Time-dependent electric and heat currents from lead $\beta$ to the
QD can be calculated from the evolution of the total number
operator of the electrons in the lead (in unit of $\hbar=1$)[39],
\begin{eqnarray}
\left({{\begin{array}{*{20}c} {J_{\beta\sigma}(t)}
\hfill \\
{Q_{\beta\sigma}(t)} \hfill \\
\end{array} }} \right) =
\frac{d}{dt}\langle \sum_k\left({{\begin{array}{*{20}c} {-e}
\hfill \\
{\varepsilon_{k\beta}(t)-E_F(t)} \hfill \\
\end{array} }}
\right) c_{k\beta\sigma}^{\dag}c_{k\beta\sigma}\rangle,
\label{currents}
\end{eqnarray}
where $E_F(t)=e[\mu_L(t)-\mu_R(t)]$ is the Fermi level. The total
electric and heat currents are $J(t)=\sum_\sigma
[J_{L\sigma}(t)-J_{R\sigma}(t)]$ and $Q(t)=\sum_\sigma
[Q_{L\sigma}(t)-Q_{R\sigma}(t)]$, respectively. In this paper we
are interested in the time averaged electric and heat currents,
which can be derived following the standard Keldysh Green function
technique as[39]:
\begin{eqnarray}
\left({{\begin{array}{*{20}c} {J}
\hfill \\
{Q} \hfill \\
\end{array} }} \right) &=& \frac{2}{h}\frac{\Gamma_L\Gamma_R}{\Gamma_L+\Gamma_R}\sum_\sigma\int{d\varepsilon}
\left({{\begin{array}{*{20}c} {-e}
\hfill \\
{\varepsilon-E_F} \hfill \\
\end{array} }}
\right)\nonumber\\ &\times& [f_L(\varepsilon)\mathrm{Im}\langle
A_{L\sigma}(\varepsilon,t)\rangle
-f_R(\varepsilon)\mathrm{Im}\langle
A_{R\sigma}(\varepsilon,t)\rangle],
\end{eqnarray}
where the line-width function $\Gamma_{L(R)}=2\pi \sum_k
V_{L(R)d}V_{L(R)d}^*\delta(\varepsilon-\varepsilon_k)$ is assumed
to be independent of the energy under wide bandwidth
approximation.
$f_{\beta}(\varepsilon)=[1+e^{(\varepsilon-\mu_\beta)/k_BT_\beta}]^{-1}$
is the Fermi distribution function of lead $\beta$ with chemical
potential $\mu_{\beta}$, temperature $T_\beta$ and Boltzmann
constant $k_B$. The quantity $A_{\beta\sigma}(\varepsilon,t)$ is
defined through the Green function as

\begin{eqnarray}
A_{\beta\sigma}(\varepsilon,t)=\int_{-\infty}^tdt_1G_\sigma^r(t,t_1)\exp[-i\varepsilon(t_1-t)-
 i\int_{t}^{t_1}\Delta_\beta(\tau)d\tau],
\end{eqnarray}
where $G_\sigma^r(t,t_1)=-i\theta(t-t')\langle \{
d_{\sigma}(t),d_{\sigma}^{\dag}(t')\} \rangle$ is the retarded
Green function, which can be obtained by the equation of motion
technique following the processes in Ref. [40]:
\begin{eqnarray}
G_\sigma^r(t,t')&=&-i\theta(t-t'))\{
(1-n_{\bar{\sigma}})\exp(-i\int_{t'}^t\varepsilon_d(\tau)d\tau-\frac{\Gamma}{2}(1-n_{\bar{\sigma}})(t-t'))\nonumber\\
&+&n_{\bar{\sigma}}\exp(-i\int_{t'}^t[\varepsilon_d(\tau)+U]d\tau-\frac{\Gamma}{2}
n_{\bar{\sigma}}(t-t'))\}.
\end{eqnarray}

It should be noted that the adopted decoupling approximation to
higher-order many-particle Green functions is sufficient for
electronic transport in the Coulomb blockade regime, but not for
higher-order tunneling process such as the subtle Kondo effect,
which is beyond the consideration of this paper. In the above
equation, $\Gamma=\Gamma_L+\Gamma_R$. Substituting the expression
of the retarded Green function into Eq. (4) and carrying out the
integrations, $A_{\beta\sigma}(\varepsilon,t)$ becomes
\begin{eqnarray}
&A&_{\beta\sigma}(\varepsilon,t)=\sum_{k,k'}J_k(\frac{\Delta_d-\Delta_\beta}{\omega})J_{k'}(\frac{\Delta_\beta-\Delta_d}{\omega})e^{i(k+k')\omega
t}\nonumber\\
&\times&\{\frac{1-n_{\bar{\sigma}}}{\varepsilon-\varepsilon_d-k'\omega+i\Gamma(1-n_{\bar{\sigma}})/2}+\frac{n_{\bar{\sigma}}}{\varepsilon-\varepsilon_d-U-k'\omega+i\Gamma
n_{\bar{\sigma}}/2}\},
\end{eqnarray}
where $J_k$ are Bessel functions of the first kind. The occupation
number $n_{\bar{\sigma}}$ in the above equations needs to be
calculated self-consistently from the equation
\begin{eqnarray}
n_{\bar{\sigma}}=\langle
\mathrm{Im}G_{\bar{\sigma}}^<(t,t)\rangle=\int
\frac{d\varepsilon}{2\pi}\sum_\beta
f_\beta(\varepsilon)\Gamma_\beta\langle
|A_{\beta\bar{\sigma}}(\varepsilon,t)|^2\rangle,
\end{eqnarray}
where $G_{\sigma}^<(t,t)$ is the lesser Green function that can be
calculated by the Keldysh equation with the help of the retarded
Green function. Once $A_{\beta\sigma}(\varepsilon,t)$ is
determined, the currents can be calculated from its time-averaged
form:
\begin{eqnarray}
\langle &A&_{\beta\sigma}(\varepsilon,t)\rangle =\sum_k
J_k^2(\frac{\Delta_d-\Delta_\beta}{\omega})\nonumber\\
&\times&[\frac{1-n_{\bar{\sigma}}}{\varepsilon-\varepsilon_d-k\omega+i\Gamma(1-n_{\bar{\sigma}})/2}+\frac{n_{\bar{\sigma}}}{\varepsilon-\varepsilon_d-U-k\omega+i\Gamma
n_{\bar{\sigma}}/2}].  \label{A(si)}
\end{eqnarray}

\section{Results}
Let us start by considering the case where the microwave field is
applied only on the dot, i.e. $\Delta_L=\Delta_R=0$ and $\Delta_d
\neq 0$. In the following numerical calculations, we set $\hbar=1$
and choose the microwave field frequency $\omega=1$ as the energy
unit. In the linear response regime ($T_L=T_R=T$) the electric
current in Eq. (3) can be written in the following compact form:
\begin{eqnarray}
J=-\frac{2e}{h}\int{d\varepsilon}[f_L(\varepsilon)-f_R(\varepsilon)]T(\varepsilon),
\end{eqnarray}
where the transmission coefficient $T(\varepsilon)$ is
\begin{eqnarray}
T(\varepsilon)&=&\frac{\Gamma_L\Gamma_R}{\Gamma_L+\Gamma_R}\sum_{k,\sigma}
J_k^2(\frac{\Delta_d}{\omega})\nonumber\\
&\times&[\frac{1-n_{\bar{\sigma}}}{\varepsilon-\varepsilon_d-k\omega+i\Gamma(1-n_{\bar{\sigma}})/2}+\frac{n_{\bar{\sigma}}}{\varepsilon-\varepsilon_d-U-k\omega+i\Gamma
n_{\bar{\sigma}}/2}]. \label{T}
\end{eqnarray}
We subsequently introduce the integrals $I_{n}(T)$ with $n=0,1,2$,
$I_n (T)=-(2/h)\int \varepsilon^n (\partial f/\partial
\varepsilon)T(\varepsilon) d\varepsilon $. The linear conductance
$G$, thermopower $S$, and the thermal conductance $\kappa$ can be
expressed as
\begin{eqnarray}
 G&=&e^2 I_0(T), \nonumber\\
S&=&-I_1(T)/[e T I_0(T)],\nonumber\\
\kappa &=& (1/T)[I_2(T)-I^2_1(T)/I_0(T)].
\end{eqnarray}
 Neglecting the thermal
conductance from the phonons as is assumed here, the figure of
merit $ZT$ is given by
\begin{eqnarray}
ZT=GS^2T/\kappa.
\end{eqnarray}
Figure 1 shows the behaviors of $G$, $\kappa$, $S$, and $ZT$
versus dot level $\varepsilon_d$ for different microwave field
strength $\Delta_d$. For $\Delta_d=0$, the linear conductance $G$
in Fig. 1(a) shows two peaks respectively positioned at
$\varepsilon_d=0$ and $-U$, exhibiting the typical Coulomb
blockade effect. As for the thermal conductance $\kappa$ in Fig.
1(b), two smooth shoulders emerge around the two main peaks. When
the microwave field is applied on the dot, photon-induced peaks
emerge in $G$ at $0\pm k\omega$ and $-U\pm k\omega$, where $k$ is
an integer. Meanwhile, the height of the main peaks is reduced.
The reason is that the microwave field excites more conduction
channels in the dot, and then the transmission probabilities
through the channels of $\varepsilon_d=0$ and $-U$ are shared and
suppressed. As for the thermal conductance $\kappa$, the height of
the main peaks first increases and then decreases with the
increase of the microwave field strength. This is because $\kappa$
is determined by both the transmission probability and the heat
transferred by each electron. Contrary to the situation in $G$,
each main peak in $\kappa$ around $\varepsilon_d=0$ or $-U$ is
contributed from more than one conduction channels, which can be
seen by comparing Fig. 1(b) with 1(a). Thus, their height may be
enhanced even if the average tunnelling probability of each
channel decreases. Here the function of the microwave field is
analogous to the changing of temperature in the leads[18-20],
which adjusts the numbers of the electron and hole around the
Fermi level to participate in transport. With increasing
temperature $T$, the average tunneling probability will be
monotonically suppressed but the heat carried by each electron
increases. As a result, the height of the peaks in $\kappa$
will exhibit a nonmonotonic dependence on $T$ [19, 20], which is
different from the behavior of the peaks in $G$. We also observe
that a peak in the thermal conductance develops at the
electron-hole symmetry point $-U/2$ due to the application of the
microwave field. This may be attributed to the photon-induced
conduction channels around this point, which can be seen from Fig.
1(a).

\begin{figure}
\centering
\includegraphics[width=10 truecm]{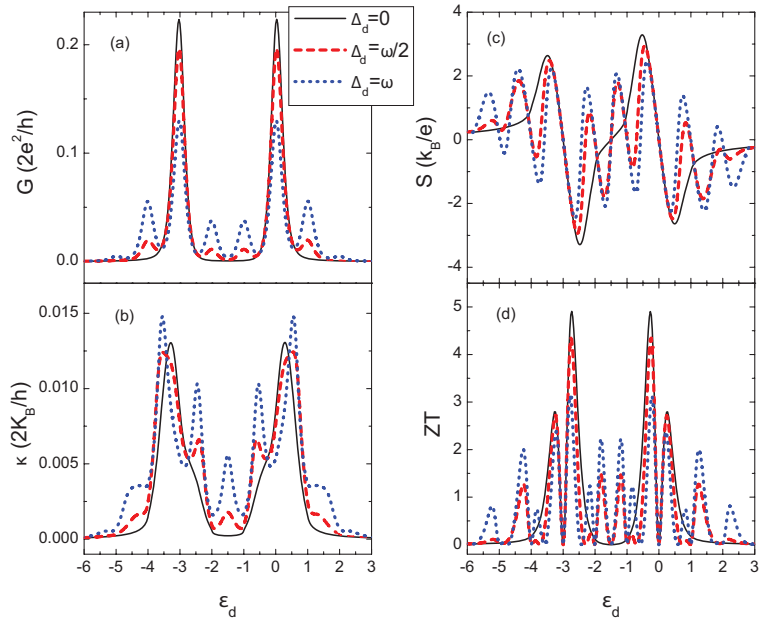}
\caption{(Color online) Electrical conductance $G$, thermal
conductance $\kappa$, thermal power $S$, and figure of merit $ZT$
as functions of dot level $\varepsilon_d$ for different microwave
field strength $\Delta_d$ with  $\Delta_L=\Delta_R=0$. Other
parameters are $U=3\omega$, $k_BT=0.1\omega$, and
$\Gamma_L=\Gamma_R=0.05\omega$. The presence of a microwave field
is seen to decrease the strength of the main peaks of the
thermoelectric figure of merit, and a set of additional peaks in
multiples of the microwave frequency emerges. \label{fig1}}
\end{figure}

The thermopower $S$ in Fig. 1(c) has three zero points
individually at $\varepsilon_d=-U$, $-U/2$ (electron-hole symmetry
point), and 0. The magnitude of $S$ oscillates between these three
points with positive and negative values, which can be understood
as follows[19,20]: In the presence of temperature difference that
induces the thermoelectric phenomenon, there are more electrons
above the chemical potential $\mu=0$ in the hotter lead, and at
the same time more holes in the cooler lead below $\mu$. When the
dot energy levels $\varepsilon_d$ or $\varepsilon_d+U$ is above
$\mu$, more electrons will transport from the left lead of higher
temperature to the right lead of lower temperature than the holes
tunnelling from the opposite direction, inducing a positive
voltage drop $\Delta V$ and consequently a negative thermopower
$S=-\Delta V/\Delta T$ ($S$ is negative in the figure as in unit
of $k_B/e$). When the energy levels are aligned to the chemical
potential or the electron-hole symmetry point, the voltage drop
induced by the electrons are cancelled by that of the holes. Thus
the thermopower is zero. When $\varepsilon_d$ or $\varepsilon_d+U$
is below the chemical potential, more electrons tunnel from right
to left and $S$ changes sign. The above explanation also holds
true when the microwave field is applied. Now a set of
sub-channels are opened by the photons, inducing more zero points
in $S$. Consequently, the applied microwave field can change the
sign of $S$, which may be used to detect the properties of the
microwave field. Fig. 1(d) shows that the magnitude of the main
peaks in $ZT$ is reduced by the microwave field with the emergence
of a set of shoulder peaks. This is because of the combined effect
of reduced linear conductance and enhanced thermal conductance.
Although the original peaks height is lowered by the microwave
field, there are several new peaks that can reach quite large
value, and may be useful for energy conversion.

Figure 2 shows the temperature-dependence of $ZT$ for different
microwave field strength $\Delta_d$. To acquire a large $ZT$
value, the dot level is fixed at -0.25$\omega$, where the main
peak in $ZT$ is centered. It is shown that the $ZT$ value first
increases and then decreases after reaching a maximum value with
the increase of the temperature, which is consistent with previous
result [18-20]. The magnitude of $ZT$ is determined by the
combined behavior of $G$, $\kappa$, and $S$. Detailed dependence
of these quantities on the temperature can be found in Refs. 19
and 20. For $k_BT\ll U$, the $ZT$ value is considerably enhanced
due to the Coulomb blockade effect, and optimal thermoelectric
efficiency can be achieved in this region. With the increase of
the temperature, the peaks in $ZT$ is suppressed as the influence
of Coulomb blockade effect on electron transport becomes less
important. In the presence of the microwave field, the line-shape
of $ZT$ is reserved but with reduced strength. This is consistent
with the above result. With the increase of the microwave field
strength, the maximum $ZT$ value shifts to lower temperature
regime.

\begin{figure}[!ht]
\centering
\includegraphics[width=10 truecm]{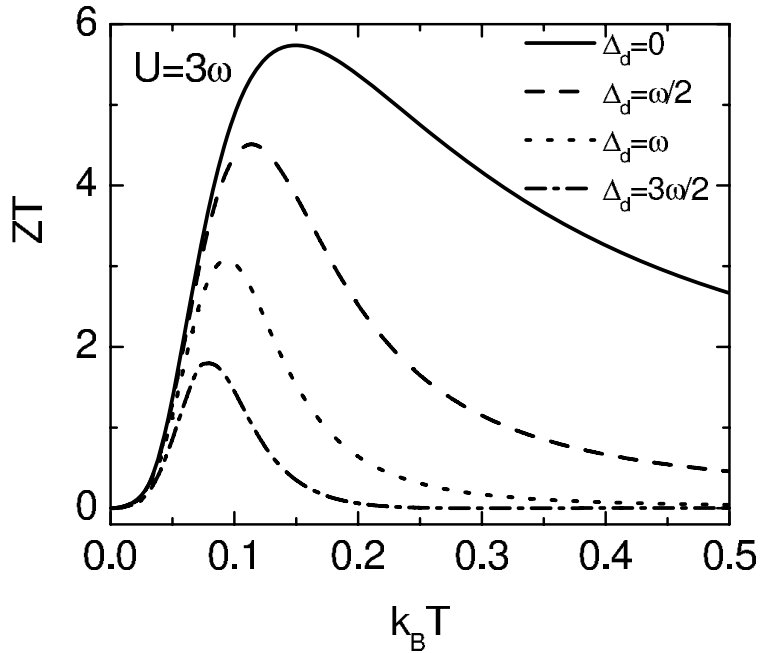}
\caption{ZT as a function of $k_BT$ for different microwave field
strength $\Delta_d=0,\omega /2, \omega, 3\omega/2$ with
$\Delta_L=\Delta_R=0$. Other parameters are as in Fig. 1. The
effect of microwave field applied only to the quantum dot is a
string reduction of ZT. \label{fig2}}
\end{figure}

When the microwave field is asymmetrically applied to the
structure, heat current and bias voltage will be generated under
the condition of zero electrical current (this condition is easily
met in thermopower experiments [1,2]). From Eqs. (3) and (8), now
the heat current is mainly determined by the difference between
the microwave field strength in the dot $\Delta_d$ and that in the
leads $\Delta_{L/R}$. We thus consider the case of a field applied
only on the leads. For comparison, we first present the results
without any microwave in Fig. 3(a-b). It shows that the heat
current always flows from the hotter lead to the cooler one under
a thermal bias $\Delta T$. It has a main resonance when the dot
level is located on the electron-hole symmetry point, and two dips
when $\varepsilon_d$ and $\varepsilon_d+U$ are aligned to the
chemical potential $\mu=0$. The reason behind this is as follows:
When $\varepsilon_d$ and $\varepsilon_d+U$ are aligned to the
chemical potential, the energy difference between the electrons
from different leads, which are tunnelling in opposite directions
through these levels, is relatively small, resulting in small heat
current. But when the dot level is at the symmetry point
$\varepsilon_d=-U/2$, electrons in the left lead tunnel to the
right lead through the level of $\varepsilon_d+U=U/2$, which is
above the chemical potential and have larger thermal energy.
Meanwhile, equal number electrons tunnel from the right lead
through the level of $-U/2$ to the left one with less thermal
energy. In other words, holes tunnel from the left lead to the
right lead through $-U/2$. The contributions from the electrons
and holes to the heat current add constructively, giving rise a
high peak. Other smaller resonances in Fig. 3(a) can be explained
by similar reasons. The bias voltage $\Delta V$ in Fig. 3(b) has
three zero points at energy levels of $\varepsilon_d=0$, $-U$ and
$-U/2$, because where equal numbers are flowing from opposite
directions. Apart from these levels, it oscillates between
negative and positive values. The reason is just the same as that
of the thermopower in Fig. 1(c). If there is microwave field
applied on the QD as shown in Fig. 3(c-d), the trends of the heat
current and electrical bias voltage are preserved, except for the
emergence of a set of smaller peaks. The heat current magnitude is
enhanced because there are more carriers participating in
transport through the photon-induced sub-channels.
\begin{figure}[!t]
\centering
\includegraphics[width=10 truecm]{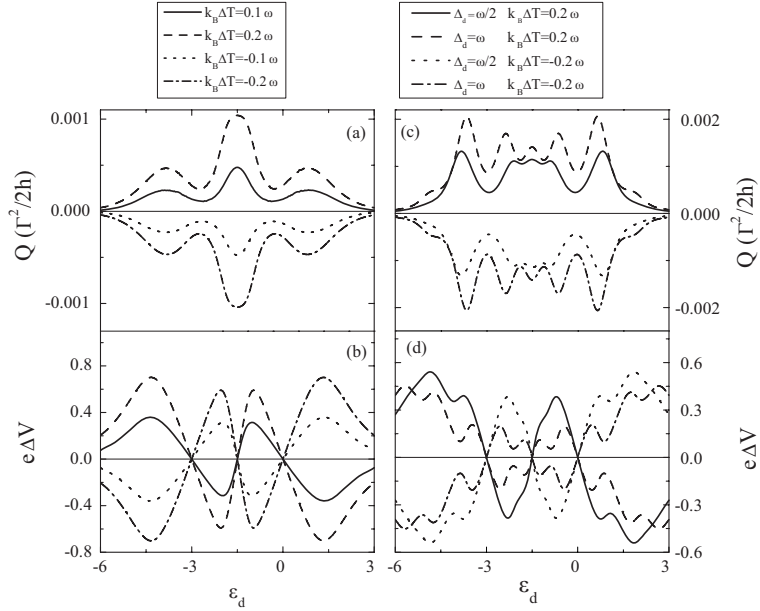}
\caption{(a) Heat current $Q$ and (b) bias voltage $\Delta V$
induced by different thermal bias $\Delta T$ as functions of dot
level $\varepsilon_d$ in the absence of microwave field. The
temperatures of the left and right leads are $T_L=T+\Delta T/2$
and $T_R=T-\Delta T/2$ with $k_BT=0.25\omega$. Other parameters
are as in Fig. 1. (c-d) same as (a-b) for different thermal bias
and in the presence of various microwave strength $\Delta_d$
(detailed in the legend), keeping $\Delta_L=\Delta_R=0$. Both the
heat current and the induced voltage exhibit a decrease of their
maximum values, and the appearance of additional microwave-induced
peaks. The heat always flows from the hot to the cold lead.
\label{fig3}}
\end{figure}

Let us now consider the situation when the microwave field is
applied only to the left lead. Fig. 4(a-b) shows the heat current
and the induced electric voltage as functions of the dot level for
various temperature differences with $\Delta_d=\Delta_R=0$ and
$\Delta_L=\omega$. Compared to the cases in Figs. 3(a-b) and
3(c-d), the magnitude of the heat current is remarkably enhanced
as the dot levels $\varepsilon_d$ and $\varepsilon_d+U$ are
slightly higher and lower than zero. For positive thermal bias
$k_B \Delta T>0$, the heat current can flow from the right cooler
(dark) lead to the left hotter (bright) lead except for
$\varepsilon_d$ is in the middle of the Coulomb blockade regime,
which is shown in Fig. 4(a). For negative thermal bias $k_B\Delta
T<0$, the heat current always flow from the right hotter (dark)
lead to the left cooler (bright) one (see the solid and dashed
lines in Fig. 4(a)). As indicated by the behavior of the electric
bias $\Delta V$ in Fig. 4(b), electron transport behavior under
both positive and negative thermal biases resembles that of
positive thermal bias case that shown in Fig. 3. Meanwhile, the
magnitude of the electric bias $\Delta V$ of positive thermal bias
is enhanced because of the photon-electrical effect. When the
microwave field is applied on the left hotter lead, the
photon-induced peaks in the bias voltage are smoothed off due to
thermal effects. The above trends of the heat current and $\Delta
V$ hold true even for zero thermal bias case.

\begin{figure}[h!]
\centering
\includegraphics[width=10truecm]{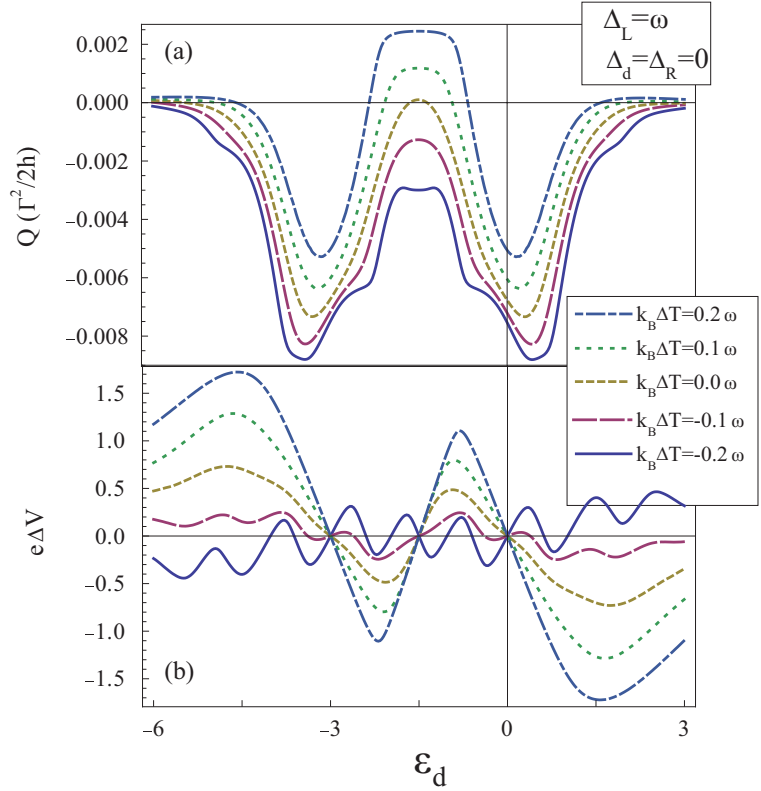}
\caption{Heat current $Q$ (a) and bias voltage $\Delta V$ (b) as
functions of dot level $\varepsilon_d$ for different thermal bias
$k_B \Delta T=-0.2 \omega,-0.1 \omega,0,0.1 \omega,0.2 \omega$. A
microwave field $\Delta_L=\omega$ is applied only to the left
lead. Other parameters are as in Fig. 3. Due to the asymmetry in
the microwave field, heat current flows from the dark to the
bright lead almost regardless of the temperature gradient, except
at the vicinity of particle-hole symmetric point
$\varepsilon_d=-U/2$.\label{fig4}}
\end{figure}

We now qualitatively discuss the impact of phonon on the figure of
merit and the heat current. At low temperature regime, we can add
to the thermal conductance $\kappa$ in Eq. (12) a phonon
contribution which is $\kappa_{ph}=3\pi^2k_B^2T/h$ [16]. Then the
magnitude of $ZT$ will be reduced correspondingly. But the trend
of $ZT$ can be unchanged [19]. In the non-linear case, phonon
peaks in the heat current may also emerge, which should be an
interesting topic. Nevertheless, recent studies indicated that the
heat current delivered by phonons can be blocked by particular
device design [19], and moreover, $\kappa_{ph}$ can be small
depending on the phonon spectrum [41,42].

\section{Discussion and summary}
The result presented in the last section is rather surprising:
although energy (via the microwave field) is pumped into the
bright lead, still heat flows from the dark to the bright lead.
The explanation for this effect we point out that the microwave
field shifts the spectral weight of the electrons in multiples of
the photon frequency, thus effectively reduces the density of
states (DOS) in some places and increases it in others. In this
case, the effective reduction of the DOS is most pronounced near
the particle-hole symmetry point, and the weight is shifted below
it. The result is electrons, which are flowing from the bright to
the dark lead, carry {\sl negative} energies, because they are
placed below the Fermi energy.

To understand this unique situation it is useful to turn to the
analogy with the non-interacting dot where $U=0$. In that case,
the dot is a simple resonant level, and consequently
$\mathrm{Im}\langle A_{\beta \sigma}(\varepsilon,t) \rangle$ is
simplified,

\begin{eqnarray} \mathrm{Im}\langle A_{\beta \sigma}(\varepsilon,t)
\rangle=\sum_k J^2_k (\frac{\Delta_\beta}{\omega})
\frac{\Gamma}{(\varepsilon-\varepsilon_d-k \omega
)^2+\Gamma^2}.\end{eqnarray} Substituting it back to Eq. (3) one
obtains for the heat current
\begin{eqnarray}
 Q&=&\frac{2}{h}\sum_k \int
\varepsilon d\varepsilon
\frac{\Gamma}{(\varepsilon-\varepsilon_d-k \omega )^2+\Gamma^2}
\left[ f_L(\varepsilon ) J^2_k(\Delta_L/\omega)-f_R(\varepsilon )
J^2_k(0) \right].
\end{eqnarray}

The Bessel functions $J_k(x)$ are decaying functions of $k$ for $x
\leq 1.434 $. Since the Fermi functions decays above the Fermi
level, the main contributions to the integral above are $k=0$ and
$k=-1$, resulting in

\begin{eqnarray} Q&\approx &\frac{2}{h} \int \varepsilon d\varepsilon
\frac{\Gamma}{(\varepsilon-\varepsilon_d )^2+\Gamma^2} \left[
f_L(\varepsilon ) J^2_0(\Delta_L/\omega)-f_R(\varepsilon )\right]
\nonumber\\ & &~~+ \frac{2}{h} \int \varepsilon d\varepsilon
\frac{\Gamma}{(\varepsilon-\varepsilon_d+\omega )^2+\Gamma^2}
f_L(\varepsilon ) J^2_1(\Delta_L/\omega) \nonumber\\ & =&
\frac{2}{h} \int \varepsilon I(\varepsilon ) d\varepsilon .
\end{eqnarray}

In the absence of the microwave field, the transmission is
centered around $\varepsilon_d$ and has a single peak, but
additional conduction channels open when the microwave field is
applied. This can be seen in Fig. 5 where $I(\varepsilon)$ (solid
line), which is proportional to the transmission function, and the
heat current integrand $\varepsilon I (\varepsilon)$ (dashed line)
plotted as functions of the integration variable $\varepsilon$,
where we have chosen the dot level $\varepsilon_d=0$ for clarity.
$I(\varepsilon)$ shows an additional peak at negative energies,
which induces electron current from the bright to dark leads. The
heat current is thus in the opposite direction. Note also that the
weight of the negative peak is larger than that of the peak near
$\varepsilon_d$, resulting in a net negative heat current.

\begin{figure}[h!]
\centering
\includegraphics[width=10 truecm]{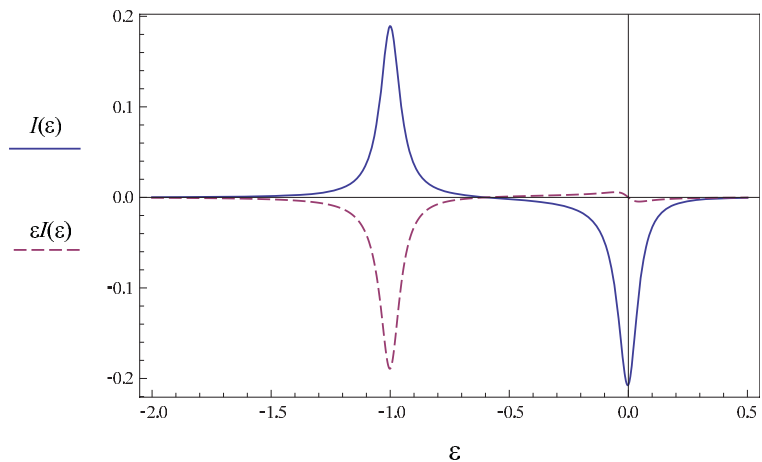}
 \caption{ (Color online) $I(\varepsilon )$ (solid line), which is proportional to
 the transmission function, and the heat current integrand $\varepsilon I (\varepsilon)$ (dashed line)
 plotted as functions of the integration variable $\varepsilon$ for $\varepsilon_d=0$. An additional peak appears in $I(\varepsilon)$, and when weighted with energy $\varepsilon I(\varepsilon)$ results in negative heat current, i.e. from dark to bright lead. } \label{Fig.5}
\end{figure}

In summary, microwave-assisted heat transport in a quantum dot is
investigated in the framework of nonequilibrium Green's function
method. It was found that the figure of merit can be suppressed by
a microwave field applied on the dot. In the completely asymmetric
case, i.e., the microwave field is applied only on one of the
leads, the heat current generally flows from the dark lead to the
bright lead, weakly depends on the direction of the thermal bias,
and can be reversed only when the dot level is close to the
electron-hole symmetry point. The results we have presented here
can be verified in experiments similar to those of Ref. [38] in
which a temperature difference is applied between the two leads as
well as a microwave field and gate voltage. We note that here we
only deal with the fixed microwave frequency case at steady state.
One can expect frequency-and time-dependent heat transport will
bring about more interesting phenomena, which are beyond the scope
of the present paper, and research in that direction is currently
underway.

\section{Acknowledge} This work was supported by NFS-China (10704011) and LNET
(2009R01).

\end{document}